\documentclass[twocolumn,english,american,prl,showpacs,preprintnumbers]{revtex4}
\usepackage[T1]{fontenc}
\usepackage[latin9]{inputenc}
\usepackage{color}
\usepackage{float}
\usepackage{units}
\usepackage{textcomp}
\usepackage{amsmath}
\usepackage{amssymb}
\usepackage{graphicx}

\makeatletter
\@ifundefined{textcolor}{}
{%
 \definecolor{BLACK}{gray}{0}
 \definecolor{WHITE}{gray}{1}
 \definecolor{RED}{rgb}{1,0,0}
 \definecolor{GREEN}{rgb}{0,1,0}
 \definecolor{BLUE}{rgb}{0,0,1}
 \definecolor{CYAN}{cmyk}{1,0,0,0}
 \definecolor{MAGENTA}{cmyk}{0,1,0,0}
 \definecolor{YELLOW}{cmyk}{0,0,1,0}
}

\usepackage{dcolumn}\usepackage{bm}

\makeatother

\usepackage{babel}
\begin{document}

\title{Theory of microwave spectroscopy of
 Andreev bound states with a Josephson junction}

\author{L.\ Bretheau$^{1,*}$, \c{C}.\ Ö.\ Girit$^{1}$, M.\ Houzet$^{2,\,3}$,
H.\ Pothier$^{1,\text{\dag}}$, D.\ Esteve$^{1}$ and C.\ Urbina$^{1}$}

\affiliation{$^{1}$Quantronics Group, Service de Physique de l'État Condensé
(CNRS, URA\ 2464), IRAMIS, CEA-Saclay, 91191 Gif-sur-Yvette, France}

\affiliation{$^{2}$Université Grenoble Alpes, INAC-SPSMS, F-38000 Grenoble, France}

\affiliation{$^{3}$CEA, INAC-SPSMS, F-38000 Grenoble, France}

\date{Received 24 June 2014; revised manuscript received 2 October 2014; published 8 October 2014 in PRB 90, 134506}

\begin{abstract}
\textcolor{black}{We present a microscopic theory for the current
through a tunnel Josephson junction coupled to a non-linear environment,
which consists of an Andreev two-level system coupled to a harmonic
oscillator. It models a recent experiment}~[Bretheau, Girit, Pothier, Esteve, and Urbina, Nature {\bf 499}, 312 (2013)] \textcolor{black}{on
photon spectroscopy of Andreev bound states in a superconducting atomic-size
contact. We find the eigenenergies and eigenstates of the environment
and derive the current through the junction due to inelastic Cooper
pair tunneling. The current-voltage characteristic reveals the transitions
between the Andreev bound states, the excitation of the harmonic mode
that hybridizes with the Andreev bound states, as well as multi-photon
processes. The calculated spectra are in fair agreement with the experimental
data.}
\end{abstract}

\pacs{74.45.+c, 73.23.-b, 74.50.+r}

\maketitle

\section{\textup{Introduction}}

\begin{figure}[H]
\begin{centering}
\includegraphics[width=1\columnwidth]{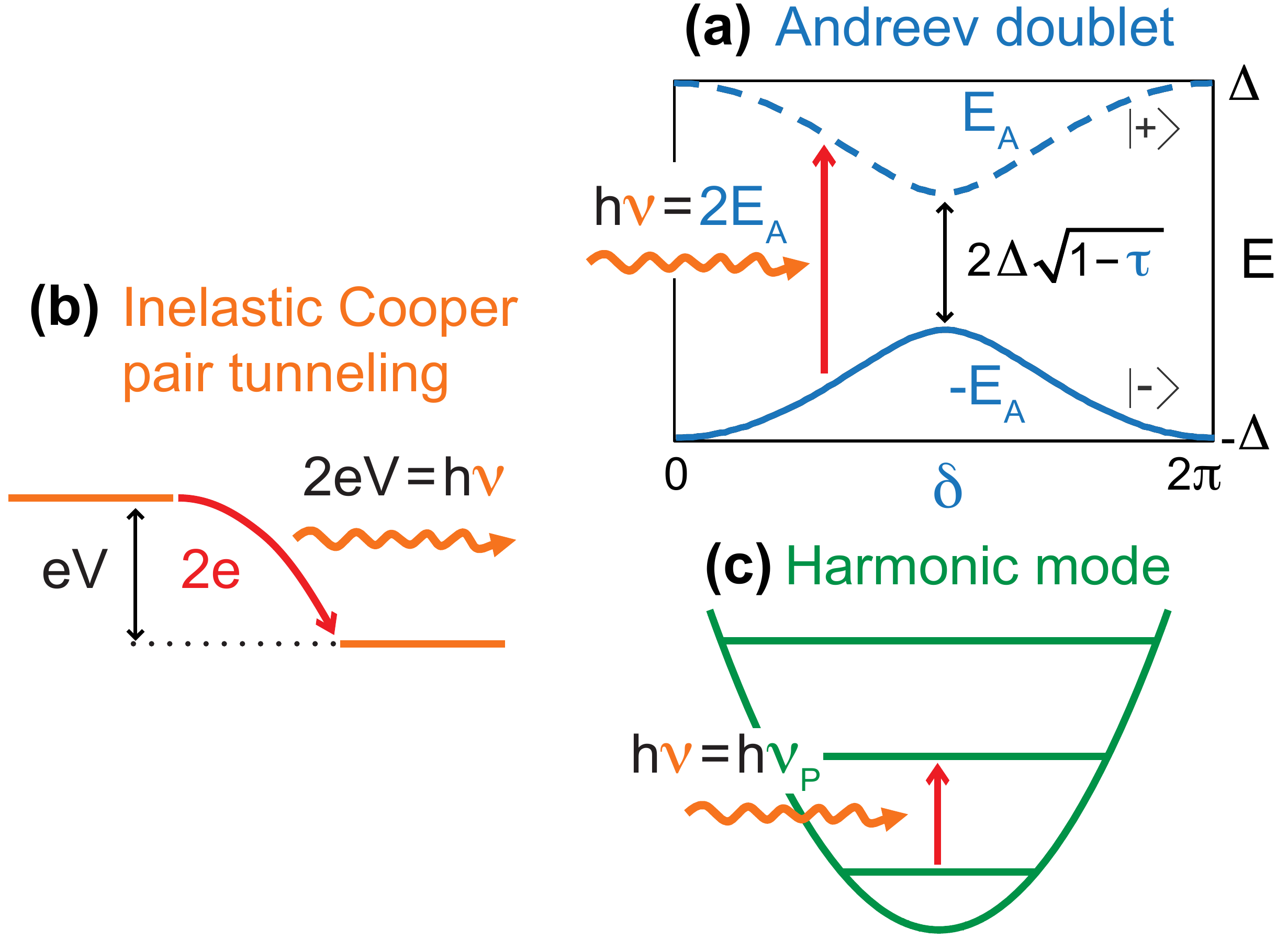}
\par\end{centering}

\caption{\label{fig:Principe spectro}(Color online) Principle of the \textcolor{black}{photon
spectroscopy of Andreev bound states}. \textbf{(a)} Phase $\left(\delta\right)$
dependence of the Andreev levels with energies $\mp E_{A}$ in a short
conduction channel of transmission $\tau.$ $\Delta$ is the superconducting
gap. \textbf{(b)}~A Cooper pair tunneling across the spectrometer
junction releases energy $2eV$ as a photon of frequency $\nu,$ which
is absorbed either in the atomic contact by exciting the Andreev transition
at energy $2E_{A}$ \textbf{(a)} or in a harmonic oscillator mode
\textbf{(c)} present in the embedding circuit of the atomic contact
(see Fig.~\ref{fig:Electrical-circuit-seen}).}
\end{figure}

The Josephson effect, predicted and observed 50 years ago \textcolor{black}{in
superconducting tunnel junctions~\cite{Nicol1960,Josephson1962,Anderson1963}},
describes the non-dissipative supercurrent that results from the coherent
coupling between superconductors separated by a thin insulating barrier.
Since then the supercurrent has been observed in many other weak links
such as point contacts, semiconducting nanowires, carbon nanotubes,
graphene sheets and thin ferromagnetic layers~\cite{Golubov2004}.
Microscopically the Josephson coupling is established through\textcolor{black}{{}
fermionic states }whose energies depend on the superconducting phase
difference $\delta$ across the weak link. For a weak link shorter
than the superconducting coherence length $\xi$, these so-called\textcolor{black}{{}
Andreev bound states (ABS)} have energies smaller than the superconducting
gap $\Delta$ and are therefore localized at the weak link over a
distance of the order of $\xi$~\cite{Kulik1970,Furusaki1991,Bagwell1992,Beenakker1992}.
In a single-channel weak link, there is only\textcolor{black}{{} one
pair of ABS, $\left|-\right\rangle $ and $\left|+\right\rangle $,}
with energies $\mp E_{A}$ (see Fig.~\ref{fig:Principe spectro}(a)),
where 
\begin{equation}
E_{A}=\Delta\sqrt{1-\tau\sin^{2}\left(\delta/2\right)}.\label{eq:ABs energy}
\end{equation}
Here, $\delta$ is the superconducting phase difference and $\tau$
is the transmission probability for electrons in the normal state.
The phase dependence gives rise to opposite supercurrents $\mp\left(1/\varphi_{0}\right)\left(\partial E_{A}/\partial\delta\right)$
in the ground and excited states (with $\varphi_{0}=\hbar/2e$ the
reduced flux quantum). This pair of states can be seen as a spin-$\nicefrac{1}{2}$
and introduces an internal degree of freedom to Josephson weak links.

At zero temperature only the lower energy state \textcolor{black}{$\left|-\right\rangle $
}is occupied. This ground state has been probed through measurements
of the current-phase relation in superconducting atomic contacts~\cite{DellaRocca2007}.
The direct observation of the higher energy state \textcolor{black}{$\left|+\right\rangle $}
has been achieved only recently~\cite{Bretheau2013,Bretheau2013b}.
In the experiment of Ref.~\cite{Bretheau2013}, a voltage-biased
Josephson junction was used as an on-chip spectrometer (see Fig.~\ref{fig:Electrical-circuit-seen}).
The dissipative subgap current through the Josephson junction is due
to inelastic tunneling of Cooper pairs, the released energy being
absorbed in the environment\textcolor{black}{{} (see Fig.~\ref{fig:Principe spectro})}.
\textcolor{black}{Current peaks are then observed at energies corresponding
to the eigenenergies of the environment. }
 \textcolor{black}{In practice, the environment was an asymmetric
SQUID formed by a superconducting atomic contact in parallel with
a second Josephson junction (see Fig.~\ref{fig:Electrical-circuit-seen}).
This environment can be modeled as a spin-$1/2$ degree of freedom
(the Andreev doublet) coupled with a harmonic mode (the plasma
mode of the Josephson junction) (see Fig.~\ref{fig:Principe spectro}).}
The goal of the present work is to reach a quantitative understanding
of the measured spectra.

The rest of the paper is organized as follows. Section~\ref{sec:The-model}
gives a Hamiltonian description of the circuit schematized in \textcolor{black}{Fig.~\ref{fig:Electrical-circuit-seen}}.
The spectrometer Hamiltonian is then treated as a perturbation, and
following previous work~\cite{Desposito2001,Zazunov2003,Zazunov2005,Bergeret2010,Romero2012,Kos2013},
the atomic contact is treated under the assumption of small phase
fluctuations. In section~\ref{sec:-spectra-:}, the resulting spin-boson
Hamiltonian of the environment is solved numerically, and the calculated
current spectra are compared with the experimental ones. Finally,
in section~\ref{sec:resolution} the Hamiltonian is solved analytically
in the Jaynes-Cummings approximation. Some perspectives are discussed
in the conclusion.

\section{Model\label{sec:The-model}}

\subsection{Setup}

\begin{figure}[h]
\begin{centering}
\includegraphics[width=1\columnwidth]{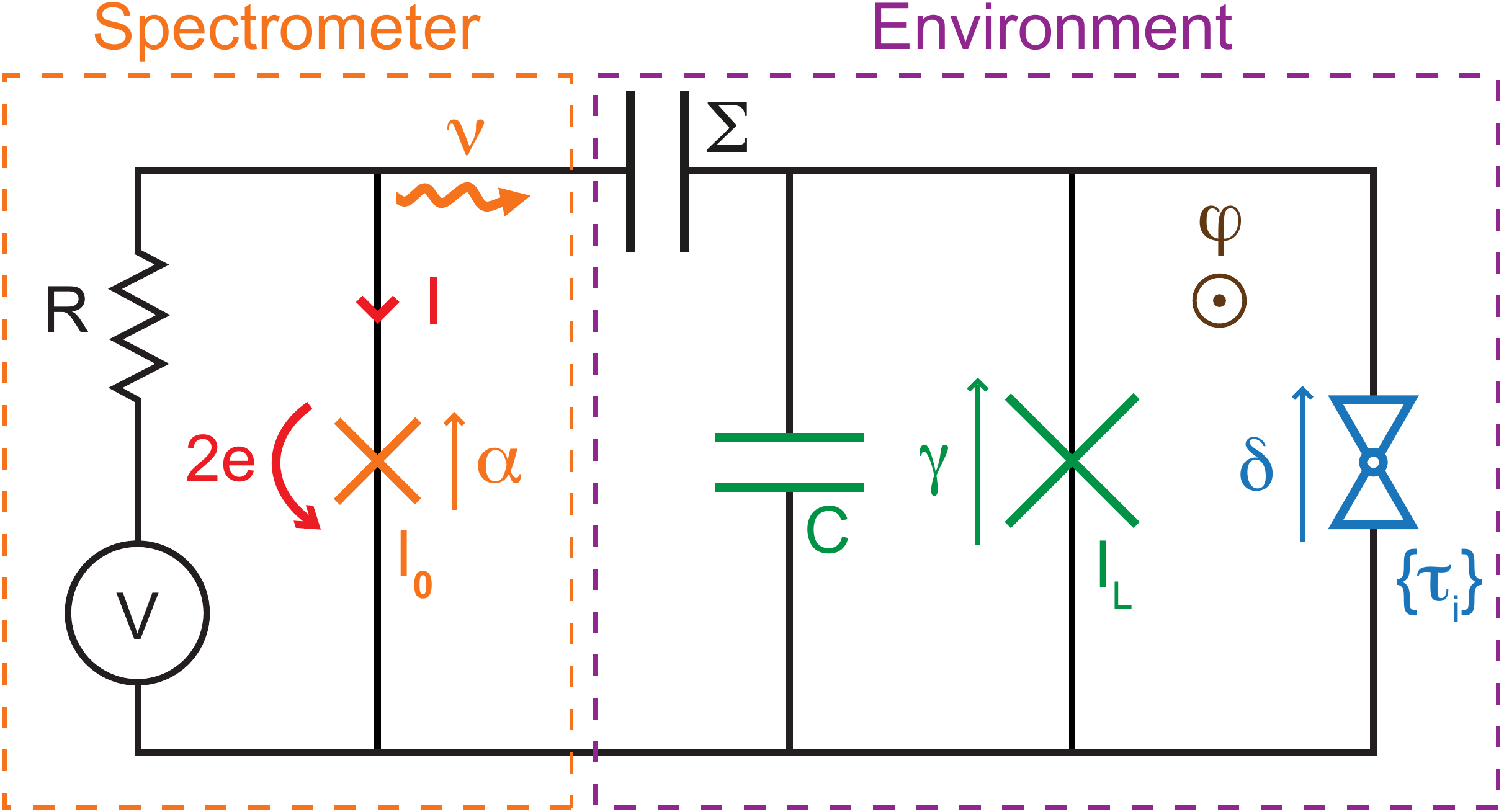}
\par\end{centering}

\caption{\label{fig:Electrical-circuit-seen}(Color online). Simplified diagram
of the experimental set-up. A voltage-biased Josephson junction (orange
cross) is used as a spectrometer. It emits microwaves in its environment,
an atomic-SQUID formed by an ancillary Josephson junction (green cross
in parallel with a capacitor $C$) and an atomic point contact (blue
triangles) of channel transmission probabilities $\left\{ \tau_{i}\right\} $.
The absorption of a photon by the environment is accompanied by the
transfer of a Cooper pair through the spectrometer. The phases $\gamma$
and $\delta$ across the SQUID junction and the atomic contact are
linked by the external reduced flux $\varphi$ threading the loop:
$\delta=\gamma+\varphi$. The phase across the spectrometer is given
by $\alpha=\gamma+2eVt$.\protect \\
}
\end{figure}

The setup used in the experiment~\cite{Bretheau2013} is shown schematically
in Fig.~\ref{fig:Electrical-circuit-seen}. \textcolor{black}{On
the left-hand side, a voltage-biased Josephson junction (critical
current $I_{0}=48$~nA) is used as a spectrometer. It is biased with
a voltage source $V$ in series with a resistor $R=2\,\textrm{k}\Omega$.
}It radiates microwaves at the Josephson frequency $\nu=2eV/h$~\cite{Note1},
which can be absorbed by its electromagnetic environment, an atomic-SQUID
formed by an atomic point contact in parallel with a second Josephson
junction. The critical current $I_{L}=1.06$~\textmu{}A of this second
junction is much larger than those of the spectrometer junction and
of a one-atom contact. An external superconducting coil is used to
apply a dc flux $\phi=\varphi_{0}\varphi$ through the SQUID loop.
The capacitance $C=280$~fF corresponds to the sum of the junctions
capacitors.\textcolor{black}{The atomic-SQUID and
the spectrometer are coupled through a capacitor $\Sigma\sim$30~pF.} 
\textcolor{black}{Whereas it behaves as an open circuit from the dc point of view and ensures that the dc voltage $V$ falls on the spectrometer, it can be considered as a short for the Josephson radiation in the explored frequency range ($\nu\gg\nu_p (C/\Sigma)^{1/2}\sim2$~GHz, with $\nu_p$ the plasma frequency of the SQUID junction).}

\subsection{Hamiltonian of the circuit}

%At high enough frequency ($\nu\gg2$~GHz),
\textcolor{black}{Neglecting the coupling capacitor and the bias resistor,} 
the circuit represented in Fig.~\ref{fig:Electrical-circuit-seen} can
be described with the Hamiltonian $H=H_{spec}+H_{SQ}$. The first
term $H_{spec}=-E_{J}\cos\left(\gamma{\scriptstyle +}2eVt\right)$
corresponds to the spectrometer with Josephson energy $E_{J}=\varphi_{0}I_{0}$.
\textcolor{black}{The voltage drop across the spectrometer junction
induces a superconducting phase difference $\alpha=\gamma+2eVt$,
}with $\gamma$ the phase at the SQUID junction~\cite{Note3}. The coupling
with the SQUID occurs through the phase operator $\gamma$. 

The Hamiltonian 
\begin{equation}
H_{SQ}=E_{C}N^{2}-E_{L}\cos\left(\gamma\right)+H_{A}\left(\delta\right)\label{eq:SQUID Hamiltonian}
\end{equation}
accounts for the SQUID. The operator $N$ is the number of Cooper
pairs that have crossed the tunnel junction; $N$ and $\gamma$ are
conjugated operators: $\left[\gamma,N\right]=i$. $E_{C}=2e^{2}/C$
is the charging energy for pairs and $E_{L}$ the Josephson energy
of the SQUID junction. The last term in Eq.~\eqref{eq:SQUID Hamiltonian}
describes the atomic contact. The phase $\delta=\varphi+\gamma$ across
the atomic contact differs from $\gamma$ by $\varphi$, the reduced
flux. For a single channel contact, and neglecting excitations which
involve quasiparticles in the continuum, the Andreev Hamiltonian is~\textcolor{black}{\cite{Zazunov2003,Zazunov2005}}
\begin{equation}
\begin{array}{c}
H_{A}=-\Delta\left(\textrm{Im}U\,\sigma_{y}+\textrm{Re}U\,\sigma_{z}\right),\\
\\
\textrm{with }U=\left[\cos(\frac{\delta}{2})+i{\scriptstyle \sqrt{1-\tau}}\sin(\frac{\delta}{2})\right]\textrm{e}^{-i{\scriptscriptstyle \sqrt{1-\tau}}\frac{\delta}{2}}
\end{array}\label{eq : Andreev Hamiltonian}
\end{equation}
where the Pauli matrices $\sigma_{x,y,z}$ act in a $2\times2$ subspace
corresponding to the two ABS. The physics of the ABS is therefore
analogous to that of a spin-$\nicefrac{1}{2}$ in a magnetic field
whose magnitude $\Delta\left|U\right|=E_{A}$ and direction depend
on the superconducting phase difference $\delta$ across the contact.
The eigenstates of $H_{A}$ are the ABS $\left|\pm\right\rangle $,
with eigenenergies $\pm E_{A}$ (see Eq.~\eqref{eq:ABs energy} and
\textcolor{black}{Fig}.~\ref{fig:Principe spectro}(a)).

\subsection{Andreev spin and plasma\textcolor{black}{{} boson}}

In the experiment~\cite{Bretheau2013}, due to the large asymmetry
$E_{L}\gg E_{A}$, the phase dynamics is essentially ruled by the
SQUID junction. The size of the phase fluctuations is determined by
the dimensionless parameter $z=\left(E_{C}/2E_{L}\right)^{1/2}\ll1$.
Therefore, following Ref.~\cite{Desposito2001}, \textcolor{black}{we
may treat the SQUID junction as a linear inductor and retain only
the lowest order coupling ($\propto\sqrt{z}$) between the SQUID junction
and the atomic contact in }Eq.~\eqref{eq:SQUID Hamiltonian}. Namely,
\begin{equation}
H_{SQ}\approx E_{C}N^{2}+E_{L}\gamma^{2}/2+H_{A}\left(\varphi\right)+\varphi_{0}\gamma C_{A}\left(\varphi\right),\label{eq:Hsq dl =0000E0 l'ordre 2}
\end{equation}
where $C_{A}=\varphi_{0}^{-1}\partial H_{A}/\partial\delta$ is the
Andreev current operator. Hence, the parallel combination of the SQUID
junction and the capacitor $C$ forms a harmonic oscillator of resonant
frequency \foreignlanguage{english}{$\nu_{p}=\sqrt{2E_{L}E_{C}}/h$}
(see \textcolor{black}{Fig}.~\ref{fig:Principe spectro}(c)). Switching
to second quantization, its Hamiltonian reads $h\nu_{p}(a^{\dagger}a+\frac{1}{2})$,
where the phase \foreignlanguage{english}{$\gamma=\sqrt{z}\left(a+a^{\dagger}\right)$}
is linked to the annihilation and creation operators $a$ and $a^{\dagger}$
of the plasma mode of the SQUID.

\begin{figure}[h]
\begin{centering}
\includegraphics[width=1\columnwidth]{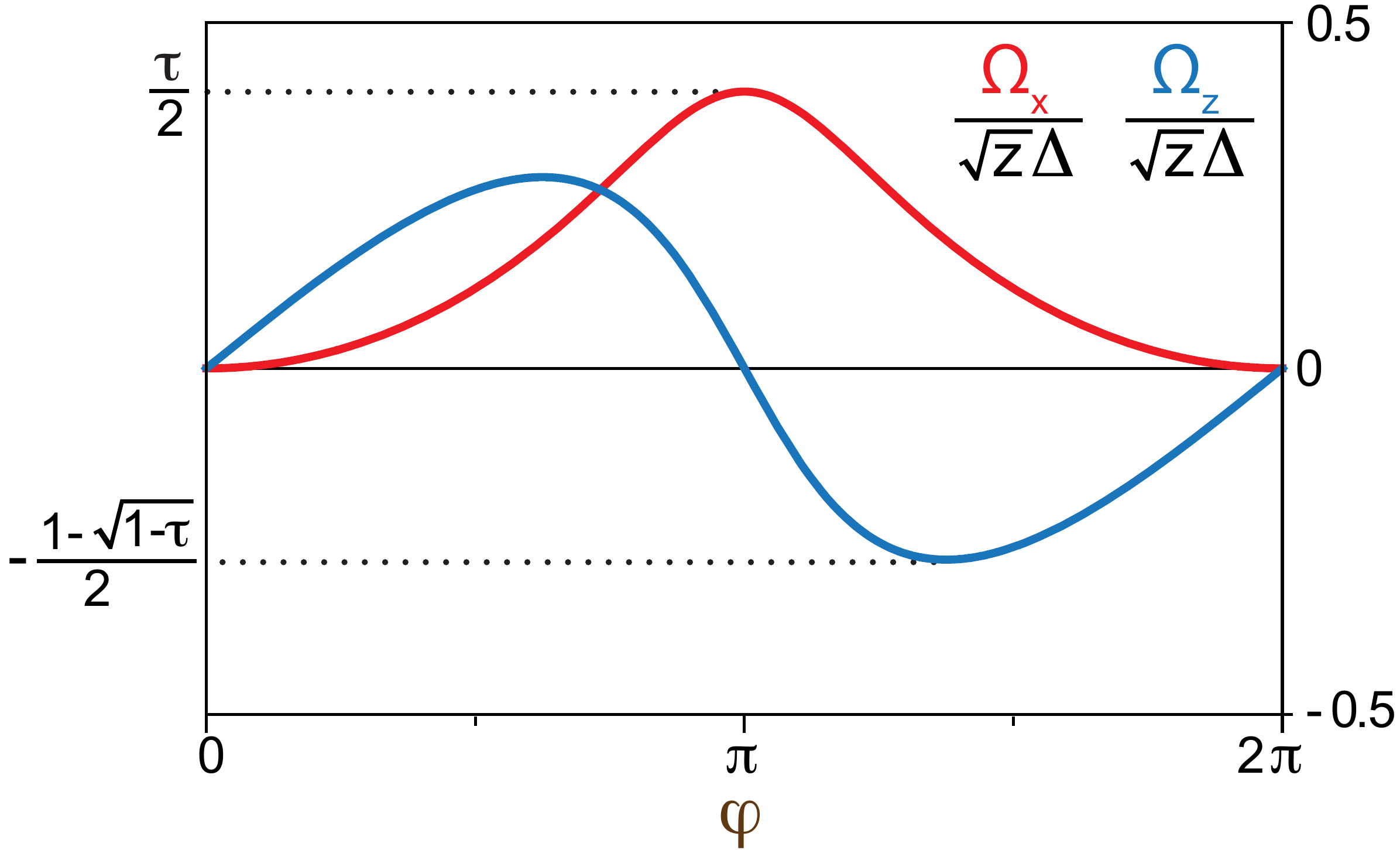}
\par\end{centering}

\caption{\label{fig:Phase-dependence-of}(Color online). \textcolor{black}{Phase
dependence of the coupling energies $\Omega_{x}$ (red) and $\Omega_{z}$
(blue) in units of $\sqrt{z}\Delta$, for $\tau=0.8$.}}
\end{figure}

In the basis of the Andreev states $\left\{ \left|-\right\rangle ,\left|+\right\rangle \right\} $,
the SQUID Hamiltonian finally reads 
\begin{equation}
H_{SQ}=h\nu_{p}\left(a^{\dagger}a+\frac{1}{2}\right)-E_{A}\sigma_{z}+\left(a+a^{\dagger}\right)\left(\Omega_{x}\sigma_{x}+\Omega_{z}\sigma_{z}\right),\label{eq:Spin-boson Hamiltonian}
\end{equation}
 where 
\begin{equation}
\Omega_{z}=\Delta\sqrt{z}\frac{\tau\sin\left(\varphi\right)}{\sqrt{1-\tau\sin^{2}\left(\frac{\varphi}{2}\right)}}\label{eq:coupling parameters}
\end{equation}
and $\Omega_{x}=\Omega_{z}\sqrt{1-\tau}\tan\left(\frac{\varphi}{2}\right)$.
\textcolor{black}{The phase-dependence of these coupling energies
is represented in }Fig.~\ref{fig:Phase-dependence-of}.

\textcolor{black}{The spin-boson Hamiltonian~\eqref{eq:Spin-boson Hamiltonian}
yields a discrete eigenspectrum which results from the hybridization
of the bosonic plasma mode in the SQUID junction and the Andreev spin-$1/2$
degree of freedom in the atomic contact} (see Fig.~\ref{fig:Excitation spectrum and probabilities with numerics}\textcolor{black}{(a)}).\textcolor{black}{{}
The term $\propto\Omega_{x}$ allows transitions between Andreev states.
It is minimum at $\varphi=0$ and maximum at $\varphi=\pi$. Note
that the flux modulation of the SQUID plasma frequency due to the
contribution of the effective inductance of the atomic contact would
appear as a higher order effect in $z$.}

\subsection{Incoherent Cooper pair current\label{sub:Incoherent-Cooper-pair}}

The global circuit Hamiltonian $H$ cannot be diagonalized analytically.
Along the lines of $P(E)$ theory for dynamical Coulomb blockade~\cite{Devoret1990,Ingold1992},
the spectrometer may be treated as a perturbation. The dc current
$I$ flowing through the spectrometer is calculated using the current
operator $I_{0}\sin(\gamma+2eVt)$ and perturbation theory up to the
second order in $E_{J}$. At zero temperature and for $V>0$, 
\begin{equation}
\frac{I\left(V\right)}{I_{0}}=\frac{\pi}{2}E_{J}\underset{k}{\sum}\left|\left\langle k\right|\textrm{e}^{i\gamma}\left|0\right\rangle \right|^{2}\delta\left((E_{k}{\scriptstyle -}E_{0}){\scriptstyle -}2eV\right),\label{eq:Current with FGR}
\end{equation}
where $\left|k\right\rangle $ are the eigenstates of $H_{SQ}$ with
energy $E_{k}$. Equation~\eqref{eq:Current with FGR} accounts for
incoherent Cooper pair tunneling through the junction's barrier at
rate $I/2e$, provided that the energy $2eV$ matches an excitation
energy of the junction environment (see \textcolor{black}{Fig}.~\ref{fig:Principe spectro}(b)).\textcolor{black}{{}
}As in $P(E)$ theory for zero temperature, we assume that, before
each tunneling event, the environment has relaxed to its ground state
$\left|0\right\rangle $.

In $P(E)$ theory~\cite{Devoret1990,Ingold1992}, the environment
is purely electromagnetic, linear and is described by an impedance
$Z(\omega)$. Equation~\eqref{eq:Current with FGR} \textcolor{black}{then
simplifies and gives the current as a function of $\textrm{Re}Z\left(\omega\right)$.
In particular, this would yield current peaks at $2eV=nh\nu_{p}$
($n$ integer), with a width related to the quality factor of the
bosonic plasma mode of frequency $\nu_{p}$.}

\textcolor{black}{Such an approach is not possible here due to the
strong non-linearity of the Andreev degree of freedom. As the model
does not include dissipation, the spin-boson Hamiltonian $H_{SQ}$
yields a discrete spectrum. Then, }Eq.~\eqref{eq:Current with FGR}
predicts infinitely sharp dc current peaks at the excitation energies
of $H_{SQ}$. A finite broadening of each of the peaks may be introduced
phenomenologically with the substitution $\delta\left(E-E_{0}\right)\rightarrow\left(\Gamma/\pi\right)/\left[\left(E-E_{0}\right)^{2}+\Gamma^{2}\right]$
in Eq.~\eqref{eq:Current with FGR}. While the model is unable to
predict the amplitude of the linewidth $\Gamma$, it should be large
enough for the incoherent Cooper pair current through the spectrometer
junction to be small, \textcolor{black}{$I\ll I_{0}$, so that perturbation
theory is valid. Near voltage $V=\epsilon_{k}/(2e)$, where }$\epsilon_{k}=E_{k}-E_{0}$,
this condition reads $(E_{J}/\Gamma)P_{k}\ll1$, with transition probability
$P_{k}=\left|\left\langle k\right|e^{i\gamma}\left|0\right\rangle \right|^{2}$.

\section{spectra calculation\label{sec:-spectra-:}}

\subsection{Numerical resolution}

\begin{figure}[h]
\begin{centering}
\includegraphics[width=1\columnwidth]{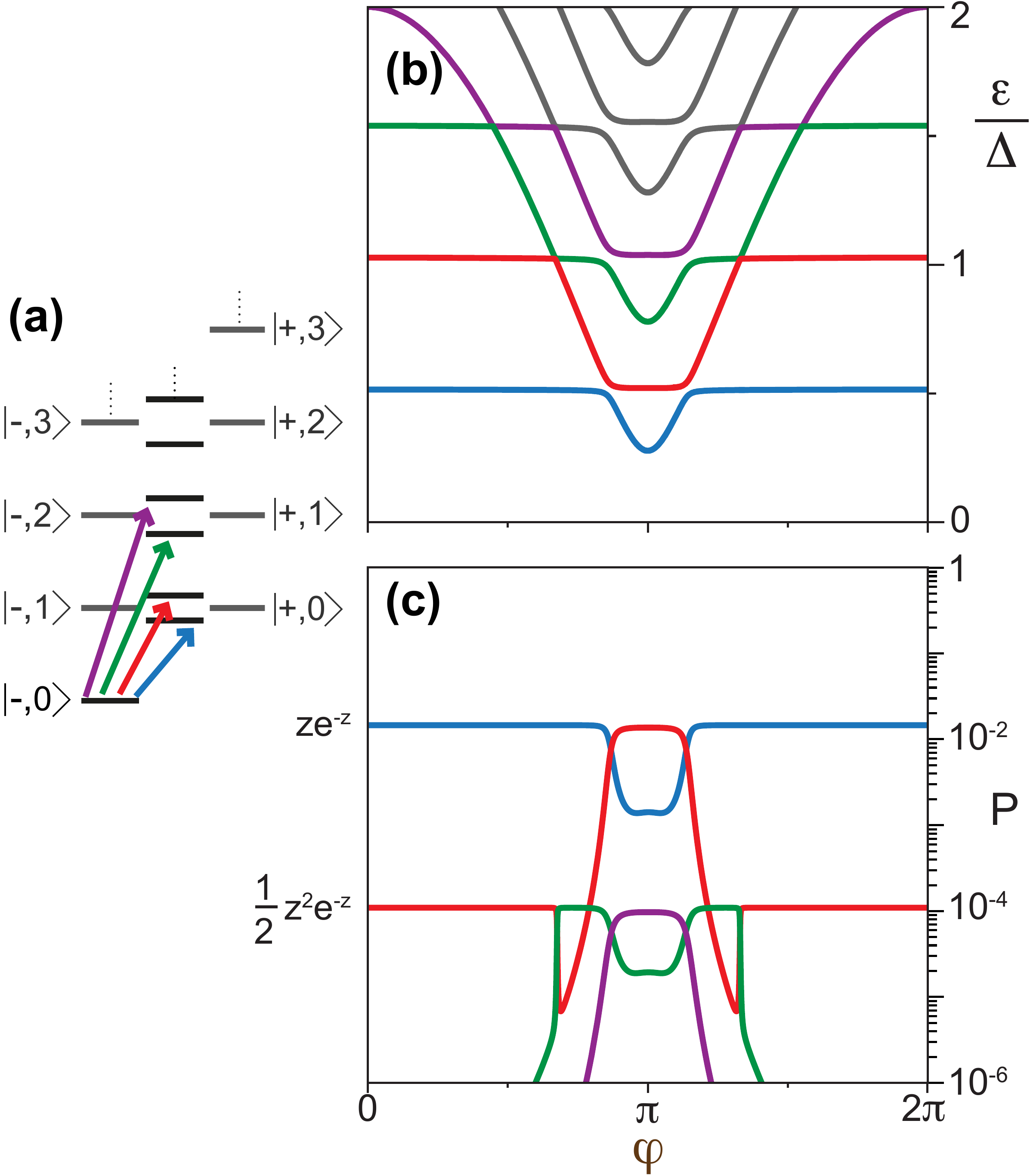}
\par\end{centering}

\caption{\label{fig:Excitation spectrum and probabilities with numerics}(Color
online). \textcolor{black}{(}\textbf{\textcolor{black}{a}}\textcolor{black}{)~Energy
spectrum diagrams }of $H_{SQ}$ for a single channel: each state is
labelled $\left|-,n\right\rangle $ or $\left|+,n\right\rangle $
for the Andreev pair in the ground ($-$) or excited ($+$) state
and $n$ photons in the plasma mode. At degeneracy $2E_{A}=h\nu_{p}$,
the plasma and Andreev modes hybridize, which leads to an avoided
crossing visible in (\textbf{b}). (\textbf{b,~c})~Excitation energies
$\epsilon$ (in units of $\Delta$) (\textbf{b}) and transition probabilities
$P$ (\textbf{c}) of the first resonances, as a function of the reduced
flux $\varphi$, for a channel of transmission $\tau=0.98$. These
lines are obtained by numerical resolution of a $14\times14$ Hamiltonian
matrix. Each color encodes a transition from the ground state towards
a different excited state, as represented in (\textbf{\textcolor{black}{a}}).}
\end{figure}

The spectrum of Hamiltonian~\eqref{eq:Spin-boson Hamiltonian} can
be obtained numerically. To do so, we write it as a matrix in the
basis $\left\{ \left|\sigma,n\right\rangle \right\} $, \textcolor{black}{where
$\sigma=\pm$ accounts for the Andreev spin-$1/2$ and $n$ is the
plasmon occupation number, }and truncate to the lower energy states.
By numerical resolution of a $14\times14$ Hamiltonian matrix $\left(n\leq6\right)$,
we have computed all the resonances of energy smaller than $2\Delta$
and the corresponding transition probabilities $P_{k}$. Figure~\ref{fig:Excitation spectrum and probabilities with numerics}
shows the excitation spectrum (b) and the transition probabilities
(c) for the transitions indicated in (a), for a channel of transmission
$\tau=0.98$.\textcolor{black}{{} In the experiment, the superconducting
gap energy is $\Delta/h\simeq43$~GHz, the charging energy is $E_{C}/h\simeq270$~MHz
and the Josephson energy of the SQUID junction is $E_{L}/h\simeq900$~GHz,
leading to }\foreignlanguage{english}{\textcolor{black}{$z\simeq0.012$}}\textcolor{black}{.
Note that $E_{L}$ is renormalized as $E_{L}=\varphi_{0}\left(I_{L}+2\varphi_{0}/L\right)$,
where $L/2=0.44$~nH is a parallel inductor (not shown in Fig.~\ref{fig:Electrical-circuit-seen})
accounting for a $1.35$~mm-long aluminum connecting wire present
in the actual geometry of the sample.}

\subsection{Incorporating line broadening}

The experimental data displayed in the right panels of \textcolor{black}{Fig.~\ref{fig:Spectra exp vs thy}}
show broad lines. One notices that the transition solely concerning
the plasma mode is already broad, which indicates that this mode is
subject to dissipation and/or dephasing; note also that voltage fluctuations
across the spectrometer junction can limit the energy resolution~\textcolor{black}{\cite{Likharev1986}}.
To account for the broadening of the energy levels of $H_{SQ}$ one
should at least include the coupling of the plasma mode to a dissipative
bath. Treating the latter as an infinite collection of harmonic oscillators,
as it is usually done, would lead to a complex spin-boson problem
without analytical solution. Here we simply introduce in Eq.~\eqref{eq:Current with FGR}
a phenomenological Lorentzian broadening, cf.~Sec.~\ref{sub:Incoherent-Cooper-pair}.
\textcolor{black}{By fitting the plasma resonance peak, we found $\Gamma/h=2$~GHz,
which we used for all the calculated lines shown in the left panels
of Fig.~\ref{fig:Spectra exp vs thy}}.

\subsection{Comparison with experiment}

\begin{figure*}[p]
\begin{centering}
\includegraphics[width=1.75\columnwidth]{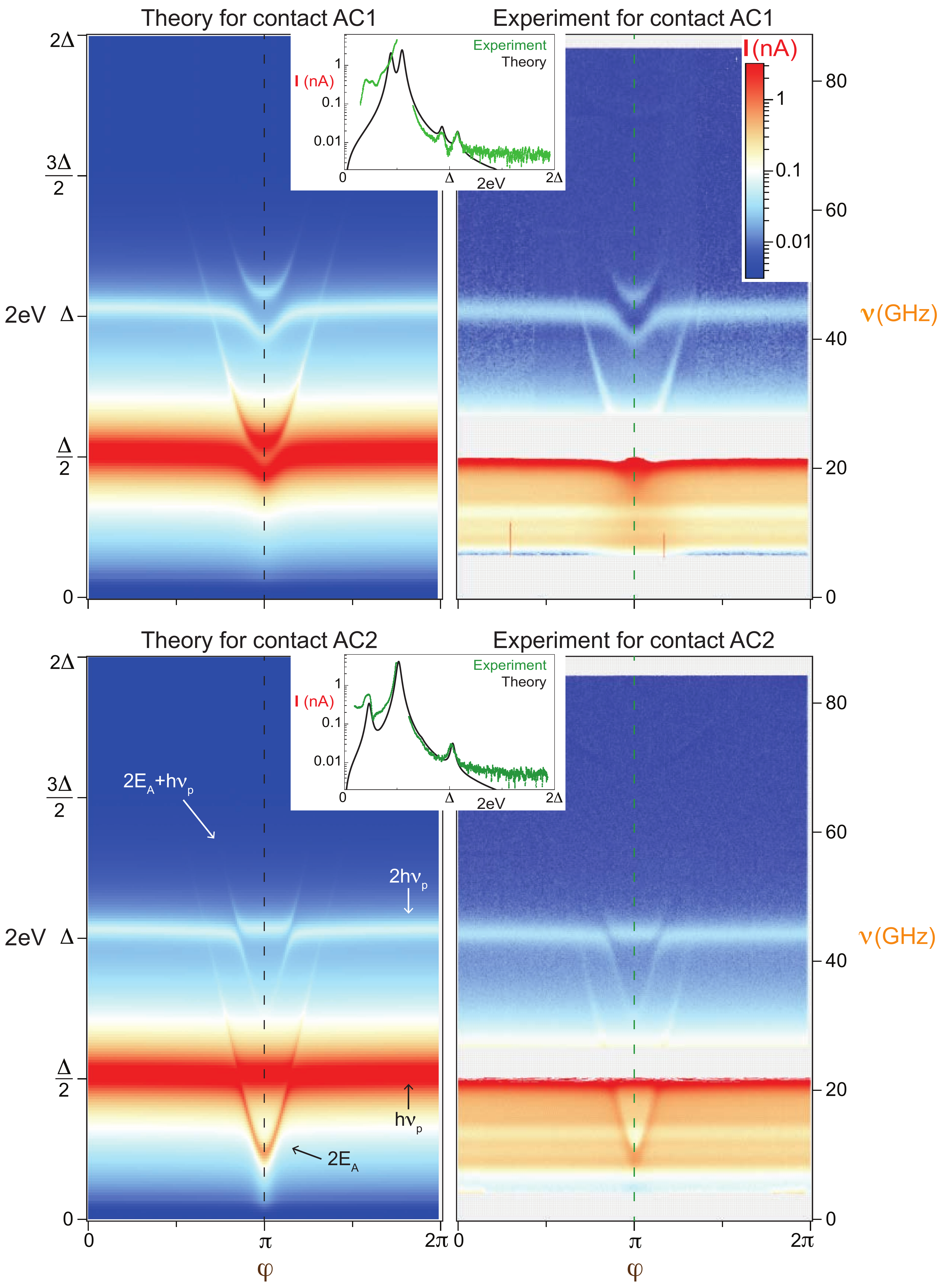}
\par\end{centering}

\caption{\label{fig:Spectra exp vs thy}(Color online). Comparison between
calculated (\textbf{left}) and experimental (\textbf{right}) spectra
$I\left(\varphi,V\right)$ for contact $AC1$ (transmissions $0.942,\,0.26$)
(\textbf{top}) and $AC2$ (transmissions $0.985,\,0.37$) (\textbf{bottom}).
\textcolor{black}{The two insets show theoretical (black lines) and experimental (green lines) $I\left(V\right)$ cuts for $\varphi=\pi$, for each contact.} \textbf{Right}:~The experimental spectra are extracted from Figure~$3$
in Ref.~\cite{Bretheau2013}. The grey regions at $\nu<4$~GHz
and around $25$~GHz are not accessible because there the biasing
of the spectrometer is unstable. \textbf{Left}:~The calculated spectra
are obtained using Eq.~\eqref{eq:Current with FGR} with a phenomenological
\textcolor{black}{damping parameter} $\Gamma/h=2$~GHz. \textcolor{black}{The
calculated currents have been multiplied by $0.6$ to mat}ch the measured
ones. \textcolor{black}{The transition probabilities and excitation
energies are }computed numerically\textcolor{black}{, using $7$ photon
levels in the plasma mode.}}
\end{figure*}

\textcolor{black}{In experiments atomic contacts have several channels.
We extend the model to contacts with multiple channels by adding }an
Andreev term~\eqref{eq : Andreev Hamiltonian} for each channel to
the Hamiltonian. This Hamiltonian describes the physics of ${\cal N}$
spins coupled to the same bosonic mode but not directly coupled to
each other. All the approximations made before are still valid~\cite{Note2}.
Figure~\ref{fig:Spectra exp vs thy} compares
the calculated spectra (left) with the experimental ones from Ref.~\cite{Bretheau2013}
(right), for two different atomic contacts $AC1$ (transmissions $0.942,\,0.26$)
and $AC2$ (transmissions $0.985,\,0.37$). The theory is obtained
\textcolor{black}{with the numerical method sketched before, still
using seven photon levels (}$28\times28$ matrix\textcolor{black}{).}

The model describes both the Andreev transitions $\left|-,0\right\rangle \rightarrow\left|+,0\right\rangle $,
of energy $2E_{A}$ (V-shaped lines), and the plasma transition $\left|-,0\right\rangle \rightarrow\left|-,1\right\rangle $,
of energy $h\nu_{p}$ (red horizontal line at $0.51\,\Delta$). It
also describes the higher harmonic transitions: $\left|-,0\right\rangle \rightarrow\left|+,1\right\rangle $,
of energy $2E_{A}+h\nu_{p}$ (replica of the Andreev transition, shifted
up by $0.51\,\Delta$) and $\left|-,0\right\rangle \rightarrow\left|-,2\right\rangle $,
of energy $2h\nu_{p}$ (white horizontal line at $1.02\,\Delta$).
These processes correspond to the tunneling of one Cooper pair and
emission of two photons. They are less probable and result in fainter
transitions as seen both in experiment and calculation. Theory also
accounts for the anti-crossings \textcolor{black}{arising} from the
coupling between the Andreev-spin and the plasma-boson.\textbf{\textcolor{black}{{}
}}Only crossings of transition lines involving the same number of
photons show significant hybridization, in good agreement with the
data. 

Extra phase-independent current at $2eV\lesssim\Delta/2$ in the experimental
data is attributed to the coupling to uncontrolled environmental modes
outside of the SQUID.

Finally, the global weakening of the signal at high $V$ is well captured
by the model. This weakening is due to the lower impedance of the
SQUID capacitance compared to its inductance at frequencies larger
than the plasma frequency. However, the agreement for the amplitude
and width of the different peaks is not quantitative. Having set the
width to 2~GHz, a correct peak amplitude is only obtained when multiplying
theory by a factor $0.6$. A rigorous treatment of dissipation is
needed. This could be achieved by adding an electromagnetic impedance
with a dissipative component in parallel with the SQUID.

\section{Andreev-plasma mode hybridization\label{sec:resolution}}

We use the rotating wave approximation, valid in the vicinity of the
degeneracy $h\nu_{p}=2E_{A}$, to obtain analytical expressions for
the excitation energies and transition rates. The Hamiltonian can
be approximated by the Jaynes-Cummings model~\cite{Jaynes1963,Shore1993}:\textcolor{black}{
\begin{equation}
H_{SQ}^{JC}=h\nu_{p}\left(a^{\dagger}a+\frac{1}{2}\right)-E_{A}\sigma_{z}+\Omega_{x}\left(a\sigma_{+}+a^{\dagger}\sigma_{-}\right)\label{eq:Jaynes Cumming Hamiltonian}
\end{equation}
}where $\sigma_{\pm}=\frac{1}{2}\left(\sigma_{x}\mp i\sigma_{y}\right)$.
Then, by block-diagonalization in the subspace $\left\{ \left|+,n\right\rangle ,\left|-,n+1\right\rangle \right\} $,
one derives the eigenstates and eigenenergies and computes the excitations
energies and transition probabilities.

\begin{figure}[h]
\begin{centering}
\includegraphics[width=1\columnwidth]{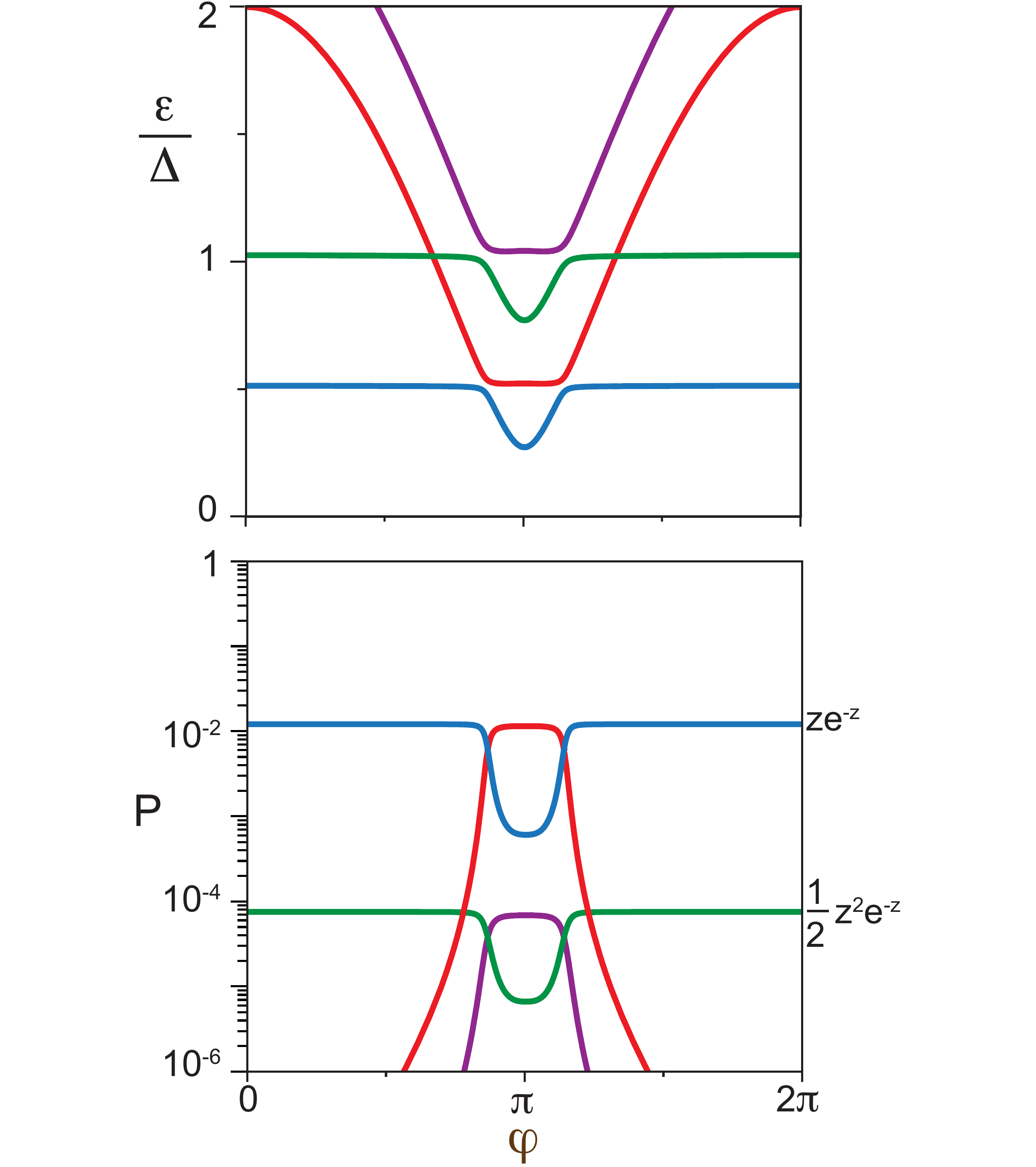}
\par\end{centering}

\caption{\label{fig:Excitation spectrum and probabilities with JC model}(Color
online). Excitation energies $\epsilon$ (in units of $\Delta$) (\textbf{top})
and transition probabilities $P$ (\textbf{bottom}) of the first resonances,
as a function of the reduced flux $\varphi$, in the Jaynes-Cummings
approximation for a channel of transmission $\tau=0.98$. Each color
encodes a transition towards a different state, as shown in Fig.~\ref{fig:Excitation spectrum and probabilities with numerics}(a).}
\end{figure}

To fourth order in $\sqrt{z}$, the current \eqref{eq:Current with FGR}
through the spectrometer displays four peaks 
\begin{equation}
\frac{I(V)}{I_{0}}=\frac{\pi}{2}E_{J}\underset{{\scriptscriptstyle \sigma=\pm;}{\scriptscriptstyle n=1,2}}{\sum}P_{n}^{\sigma}\delta\left(2eV-\epsilon_{n}^{\sigma}\right)\label{eq:I(V) JC}
\end{equation}
at the bias voltages matching the excitation energies (see top panel
of Fig.~\ref{fig:Excitation spectrum and probabilities with JC model}):
\begin{equation}
\epsilon_{n}^{\pm}=E_{A}+\left(n-\frac{1}{2}\right)h\nu_{p}\pm W_{n}\label{eq:Resonances JC}
\end{equation}
with\textcolor{black}{{} $W_{n}=\sqrt{(E_{A}-\frac{1}{2}h\nu_{p})^{2}+n\Omega_{x}^{2}}$.
T}he amplitudes of the current peaks are proportional to the transition
probabilities (see bottom panel of Fig.~\ref{fig:Excitation spectrum and probabilities with JC model}):
\begin{equation}
\left\{ \begin{array}{c}
P_{n}^{-}=\frac{1}{n!}z^{n}\textrm{e}^{-z}\cos^{2}\left(\frac{\theta_{n}}{2}\right)\\
P_{n}^{+}=\frac{1}{n!}z^{n}\textrm{e}^{-z}\sin^{2}\left(\frac{\theta_{n}}{2}\right)
\end{array}\right.\label{eq:Weight JC}
\end{equation}
\textcolor{black}{with $\theta_{n}=\arctan\left(n\Omega_{x}/W_{0}\right)$.}

These peaks correspond to excitations towards composite states, resulting
from the hybridization of the plasma and Andreev modes (see Fig.~\ref{fig:Excitation spectrum and probabilities with numerics}\textcolor{black}{(a)}).
The first two resonances in Eq.~\eqref{eq:I(V) JC} correspond to
the excitation with a single photon of the hybridized Andreev ($2E_{A}$)
and plasma mode ($h\nu_{p}$); the last two resonances correspond
to the excitation of higher harmonic modes at $2h\nu_{p}$ and $h\nu_{p}+2E_{A}$.
These two-photon processes are possible because the spectrometer,
a Josephson tunnel junction, is a nonlinear emitter. They correspond
to the tunneling of one Cooper pair and emission of two photons. Note
that far from degeneracy, using perturbation theory with $z\ll1$,
one finds for the plasma resonance and its harmonic the transition
probabilities $z\textrm{e}^{-z}$ and $\frac{1}{2}z^{2}\textrm{e}^{-z}$.
These amplitude are consistent with the Poisson distribution found
at arbitrary $z$ in the $P(E)$ derivation with a bosonic mode~\cite{Ingold1992}.

It is worth mentioning that the Jaynes-Cummings model gives an excellent
description, even far from degeneracy when\textcolor{black}{{} $\left|W_{0}\right|\gg h\nu_{p}$.}
There, the error made in energy scales as $\Omega_{x}/\left(2E_{A}+h\nu_{p}\right)$,
and is in general negligible. However, this model fails to predict
the anti-crossing between the second harmonic of the plasma resonance
and the Andreev resonance at $2h\nu_{p}=2E_{A}$ (red and green lines
in \textcolor{black}{Fig.~}\ref{fig:Excitation spectrum and probabilities with JC model}).
In practice this correction is one order of magnitude smaller than
the anti-crossing at $h\nu_{p}=2E_{A}$ and is not seen in the experiment~\cite{Bretheau2013}.
Note however that the derivation in the case of multiple channels
is much more involved when more than one Andreev transition crosses
the plasma resonance.

\section{Conclusions and perspectives}

In conclusion, we have treated inelastic Copper pair tunneling through
the emitter junction using perturbation theory in its coupling with
the environment, and calculated the environmental transition energies
and probabilities by diagonalization of the SQUID Hamiltonian. This
Hamiltonian contains a spin-like term describing the atomic contact
and which is coupled to a harmonic term for the plasma mode of the
SQUID junction. The calculated spectra are in fair agreement with
the experimental ones. However this theory could benefit from several
extensions.

First of all, the atomic contact Hamiltonian was restricted to a two-level
system Hamiltonian, by neglecting excitations which involve quasiparticles
in the continuum. In particular, quasiparticle trapping, which was
measured in a similar experiment~\cite{Bretheau2013b}, leaves the
Andreev doublet in a long-lived ``odd'' state for which both the
Andreev states are either occupied or empty~\cite{Chtchelkatchev2003},
with time scales in the millisecond range~\cite{Zgirski2011}. It limits the
Andreev transition rate and therefore reduces the corresponding dc
current. On the other hand, transitions from the Andreev bound states
to the continuum were measured on the same contacts using another
detection method~\cite{Bretheau2013b}. They do not contribute to
the dc current and were not seen in the experiment described in Ref.~\cite{Bretheau2013}
due to the long lifetime of the odd states. To deal with all the Andreev
excitations on the same footing, one must consider the full Hamiltonian
of the atomic contact, as done in recent works~\cite{Kos2013,Olivares2014}.

Second, the description used here for the environment eigenstates
and for the spectrometer is far from being complete. As discussed
above, this is a difficult problem for two reasons: (i) the coupling
of a two-level system to a dissipative resonator is not known analytically, and
(ii) the $I(V)$ characteristic of a Josephson junction coupled to
a resonator is not generically known when the perturbative expansion
in $E_{J}$ is not valid, despite recent progress along this direction~\cite{Gramich2013}.
Although the second order expansion is valid for the data discussed
here since the Cooper pair current is small ($I\ll I_{0}$), higher
order contributions corresponding to the transfer of more than one
Cooper pair emitting several photons were observed in \textcolor{black}{other}\textcolor{black}{{}
}experiments~\cite{Bretheau2013a,Bretheau2013b}.

Finally, it was also assumed in the model that the environment is
in its ground state each time a Cooper pair tunnels. This requires
the relaxation time of the excited environment states to be shorter
than the inverse tunnel rates. When this is not the case, the environment
modifies the tunneling process. For the harmonic oscillator bosonic
mode, this could lead to stimulated emission and lasing. For the Andreev
two-level system, this could saturate absorption and reduce the Cooper
pair current through the spectrometer junction. A regime in which
the Andreev two-level system would coherently exchange an excitation
with the Josephson junction can also be envisioned.

\section{Acknowledgments}

We thank Sebastian Bergeret, Juan Carlos Cuevas, Andrew C. Doherty,
Alfredo Levy Yeyati, Fabien Portier, Yuli Nazarov and Vitaly Shumeiko
for enlightening discussions. We gratefully acknowledge help from
other members of the Quantronics group. Work partially financed by
ANR through projects ANR-09-BLAN-0199-01 DOC-FLUC, ANR-12-BS04-0016 MASH and ANR-11-JS04-003-01
GLASNOST. \textcolor{black}{The research leading to these results
has received funding from the People Programme (Marie Curie Actions)
of the European Union\textquoteright{}s Seventh Framework Programme
(FP7/2007-2013) under REA grant agreement n\textdegree{} PIIF-GA-2011-298415.}

\medskip{}

\noindent {\small{$*$ Present adress: Laboratoire Pierre Aigrain,
Ecole Normale Supérieure, CNRS (UMR 8551), 75231 Paris Cedex 05, France}}{\small \par}

\noindent {\small{$\text{\dag}$ hugues.pothier@cea.fr}}{\small \par}

%\bibliographystyle{h-physrev}
%\bibliography{BiblioTheoryABSspectro}

\end{document}